\documentclass[]{mn2e}
\usepackage{epsfig}

\title[VLT \& Spitzer spectroscopy of a massive YSO in the LMC]
{ESO-VLT and Spitzer spectroscopy of IRAS\,05328$-$6827: a massive young
stellar object in the Large Magellanic Cloud}
\author[Jacco Th. van Loon et al.]{Jacco Th. van Loon$^1$,
J.M. Oliveira$^1$,
P.R. Wood$^2$,
A.A. Zijlstra$^3$,
G.C. Sloan$^4$,
\newauthor
M. Matsuura$^3$,
P.A. Whitelock$^5$,
M.A.T.\ Groenewegen$^6$,
J.A.D.L. Blommaert$^6$,
\newauthor
M.-R.L. Cioni$^7$,
S. Hony$^6$,
C. Loup$^8$,
L.B.F.M. Waters$^{9,6}$
\\
$^{1}$Astrophysics Group, School of Physical \& Geographical Sciences, Keele
      University, Staffordshire ST5 5BG, UK\\
$^{2}$Research School of Astronomy and Astrophysics, Australian National
      University, Cotter Road, Weston Creek, ACT 2611, Australia\\
$^{3}$School of Physics and Astronomy, University of Manchester, Sackville
      Street, P.O.Box 88, Manchester M60 1QD, UK\\
$^{4}$Department of Astronomy, Cornell University, 108 Space Sciences
      Building, Ithaca NY 14853-6801, USA\\
$^{5}$South African Astronomical Observatory, P.O. Box 9, 7935 Observatory,
      South Africa\\
$^{6}$Instituut voor Sterrenkunde, Celestijnenlaan 200B, B-3001 Leuven,
      Belgium\\
$^{7}$Institute for Astronomy, University of Edinburgh, Royal Observatory,
      Blackford Hill, Edinburgh EH9 3HJ, UK\\
$^{8}$Institut d'Astrophysique de Paris, CNRS, 98bis Boulevard Arago, 75014
      Paris, France\\
$^{9}$Astronomical Institute, University of Amsterdam, Kruislaan 403, 1098 SJ
      Amsterdam, The Netherlands}
\date{Submitted 2005}
\pagerange{\pageref{firstpage}--\pageref{lastpage}}
\pubyear{2005}
\begin{document}
\maketitle
\label{firstpage}
\begin{abstract}
We present the first thermal-infrared spectra of an extra-galactic Young
Stellar Object (YSO), IRAS\,05328$-$6827 in the H\,{\sc ii} region
LHA\,120-N\,148 in the Large Magellanic Cloud. The observed and modelled
spectral energy distribution reveals a massive YSO, $M\sim20$ M$_\odot$, which
is heavily-embedded and probably still accreting. The reduced dust content as
a consequence of the lower metallicity of the LMC allows a unique view into
this object, and together with a high C/O ratio may be responsible for the
observed low abundance of water ice and relatively high abundances of methanol
and CO$_2$ ices.
\end{abstract}

\section{Introduction}

Massive stars play an important r\^{o}le in galactic evolution. They deposit
chemically enriched matter, momentum and ionizing photons into the
interstellar medium (ISM), and their demise as supernovae may trigger or
quench star formation in their vicinity. Because of their rapid evolution it
is difficult to study the formation of massive stars. Until the protostar has
become hot enough to create an (ultra-compact) H\,{\sc ii} region it remains
hidden from view inside its dust cocoon (e.g., Walmsley 1995; Garay \& Lizano
1999). Studying the formation of massive stars in nearby, metal-poor
environments has the advantage of reduced attenuation by dust, as the
dust-to-gas ratio is a function of the metal abundance. An important
scientific reason for studying massive star formation at low metallicity is
that it may be more representative of the early evolution of the Universe.

Surveys for extra-galactic pre-main sequence stars have so far concentrated on
prominent H\,{\sc ii} regions in the Magellanic Clouds. Optical and near-IR
images reveal young stellar objects (YSOs) that have just emerged from their
dust cocoon or are in the process of doing so (Rubio et al.\ 1998; Walborn et
al.\ 1999; Brandner et al.\ 2001; Romaniello, Robberto \& Panagia 2004).
Embedded YSOs have been found by their dust emission at thermal-IR wavelengths
or via maser emission at radio wavelengths (Whiteoak et al.\ 1983; Epchtein,
Braz \& S\`{e}vre 1984; van Loon \& Zijlstra 2001; Maercker \& Burton 2005).

The mid-IR point source IRAS\,05328$-$6827 is seen in the direction of the
Large Magellanic Cloud (LMC), at RA$=5^{\rm h}32^{\rm m}38.59^{\rm s}$ and
Dec$=-68^\circ25^\prime22.2^{\prime\prime}$ (J2000). On the basis of its
near-IR colours and MSX 8.28 $\mu$m flux density it was selected as a target
for our thermal-IR spectroscopic programme with the ESO-VLT and Spitzer Space
Telescope of evolved stars in the Magellanic Clouds. Here we present evidence
for it to be a massive YSO instead. We present and analyse the spectra and the
spectral energy distribution, the first detailed investigation of an
extra-galactic YSO.

\section{Observations}

\subsection{Spitzer Space Telescope spectroscopy}

The InfraRed Spectrograph (IRS; Houck et al.\ 2004) on-board the Spitzer Space
Telescope (Werner et al.\ 2004) was used on 15 February 2005 to take a
low-resolution spectrum ($\lambda/\Delta\lambda\sim100$) of IRAS\,05328$-$6827
between 5.2 and $\sim36$ $\mu$m, as part of Cycle 1 programme \#3505. The star
was acquired on the slit after peak-up in the red (22 $\mu$m) on the target
itself. The spectra were taken in staring mode, placing the target on
different positions on the slit to facilitate removal of sky emission. The
spectral segments received a total integration of 28, 28, 60 and 420 seconds
for respectively SL Order 2 (5.2--8.7 $\mu$m), SL Order 1 (7.4--14.5 $\mu$m),
LL Order 2 (14.0--21.3 $\mu$m) and LL Order 1 (19.5--38.0 $\mu$m). The
spectral segments were combined by making corrections for pointing-induced
throughput errors, normalizing to the best centered segment. The accuracy of
the overall spectral shape is a few per cent. The reduction, extraction and
calibration of the spectrum followed closely the procedures described in Sloan
et al.\ (2004).


\subsection{ESO-VLT 3--4 $\mu$m spectroscopy}

The ISAAC instrument on the European Southern Observatory (ESO) Very Large
Telescope (VLT), Chile, was used on 7 December 2003 to obtain a long-slit
spectrum between 2.85 and 4.15 $\mu$m. The resolving power of
$\lambda/\Delta\lambda\sim600$ was set by the $\sim0.6^{\prime\prime}$ seeing
rather than the $2^{\prime\prime}$ slit width.

The thermal-IR technique of chopping and nodding was used, with a throw of
$10^{\prime\prime}$ and jittering by small amounts, to remove the background.
The spectra were extracted using an optimal extraction algorithm. An internal
Xe+Ar lamp was used for wavelength calibration The relative spectral response
was calibrated by dividing by the spectrum of the B-type standard star
HIP\,020020. This removed most of the telluric absorption lines but introduced
artificial emission features due to the photospheric lines of HIP\,020020. The
spectrum of IRAS\,05328$-$6827 was therefore multiplied by a hot blackbody
continuum with Gausian-shaped absorption lines of Br$\alpha$ 4.052, Pf$\gamma$
3.741, Pf$\delta$ 3.297, Pf$\epsilon$ 3.039 and Pf$\zeta$ 2.873 $\mu$m scaled
to match those in HIP\,020020.

\subsection{Photometry}

$HKL$ photometry of IRAS\,05328$-$6827 obtained at the South African
Astronomical Observatory (SAAO) was published by van Loon (2000). $JHKL$
imaging photometry was obtained by us (PRW) at Mount Stromlo and Siding
Springs Observatory (MSSSO) using the ANU 2.3m telescope with CASPIR (McGregor
1994; Wood, Habing \& McGregor 1998). $JHK_{\rm s}$ photometry is available
from the 2-Micron All-Sky Survey (2MASS; Cohen, Wheaton \& Megeath 2003).
Additional L$^\prime$-band photometry is derived from our ISAAC spectroscopy
acquisition images.

An 8.28 $\mu$m flux density is taken from version 2.3 of the Mid-course Space
eXperiment (MSX) Point Source Catalogue (Egan \& Price 1996; Egan, Van Dyk \&
Price 2001). We collected scans from the IRAS data server
(http://www.astro.rug.nl/IRAS-Server/) to measure 12 and 25 $\mu$m flux
densities, using the Groningen Image Processing SYstem (GIPSY) software with
the {\sc scanaid} tool.

The IRS red peak-up image provided photometry at 22 $\mu$m (between 18.5 and
26 $\mu$m). The individual, shifted integrations were combined to remove bad
pixels, and aperture photometry was obtained assuming a calibration of 1.16
data numbers $\mu$Jy$^{-1}$ for a 4-pixel aperture radius.

All the available photometry is listed in Table 1. No colour corrections were
applied to the photometry.

%
%
\begin{table}
\caption[]{Photometry of IRAS\,05328$-$6827 (flux densities from MSX, IRAS \&
IRS are given in Jy).}
\begin{tabular}{llllllr}
\hline\hline
Ban\rlap{d}     &
System          &
$\lambda_0$     &
$\Delta\lambda$ &
mag             &
$\sigma$        &
date            \\
\hline
J               &
2MASS           &
1.235           &
0.162           &
\llap{1}6.65    &
0.20            &
06 02 2002      \\
J               &
CASPIR          &
1.3             &
0.12            &
\llap{1}6.05    &
0.10            &
27 01 2005      \\
J               &
CASPIR          &
1.3             &
0.12            &
\llap{1}6.53    &
0.12            &
13 03 2005      \\
H               &
SAAO            &
1.65            &
0.15            &
\llap{1}4.02    &
0.05            &
11 12 1997      \\
H               &
2MASS           &
1.662           &
0.251           &
\llap{1}4.24    &
0.06            &
06 02 2002      \\
H               &
CASPIR          &
1.7             &
0.12            &
\llap{1}4.04    &
0.03            &
27 01 2005      \\
H               &
CASPIR          &
1.7             &
0.12            &
\llap{1}4.21    &
0.03            &
13 03 2005      \\
K$_{\rm s}$     &
2MASS           &
2.159           &
0.262           &
\llap{1}1.98    &
0.03            &
06 02 2002      \\
K               &
SAAO            &
2.2             &
0.19            &
\llap{1}1.90    &
0.02            &
11 12 1997      \\
K               &
CASPIR          &
2.22            &
0.2             &
\llap{1}1.93    &
0.02            &
19 11 1999      \\
K               &
CASPIR          &
2.22            &
0.2             &
\llap{1}1.78    &
0.01            &
16 02 2000      \\
K               &
CASPIR          &
2.22            &
0.2             &
\llap{1}1.72    &
0.01            &
27 01 2005      \\
K               &
CASPIR          &
2.22            &
0.2             &
\llap{1}1.81    &
0.01            &
13 03 2005      \\
L               &
SAAO            &
3.5             &
0.27            &
9.13            &
0.07            &
11 12 1997      \\
L               &
CASPIR          &
3.59            &
0.07            &
8.89            &
0.04            &
27 01 2005      \\
L               &
CASPIR          &
3.59            &
0.07            &
8.99            &
0.05            &
13 03 2005      \\
L$^\prime$      &
ESO             &
3.78            &
0.29            &
8.65            &
0.10            &
07 12 2003      \\
A               &
MSX             &
8.28            &
2.0             &
0.25\rlap{7}    &
0.01\rlap{2}    &
                \\
12              &
IRAS            &
\llap{1}2       &
3.5             &
0.4             &
0.05            &
                \\
red             &
IRS             &
\llap{2}2       &
3.7             &
0.83            &
0.08            &
15 02 2005      \\
25              &
IRAS            &
\llap{2}5       &
5.6             &
1.0             &
0.15            &
                \\
\hline
\end{tabular}
\end{table}

\section{Results}

\subsection{The environment of IRAS\,05328$-$6827}

%
%
\begin{figure}
\centerline{\psfig{figure=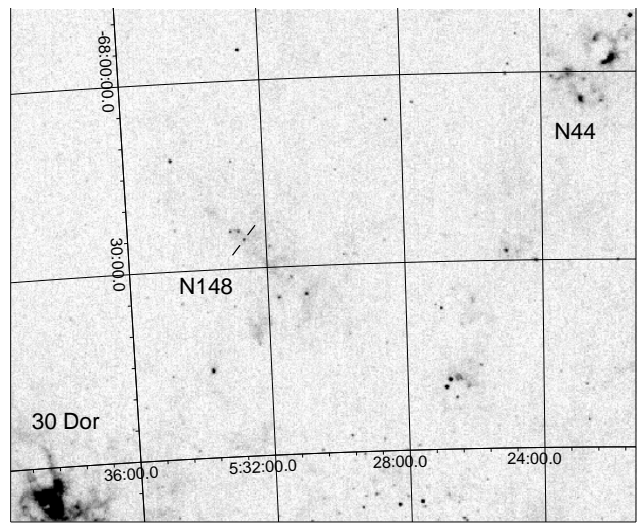,width=84mm}}
\caption[]{MSX image at 8.28 $\mu$m showing the location in the LMC of the
star forming region LHA\,120-N\,148.}
\end{figure}

%
%
\begin{figure*}
\centerline{\psfig{figure=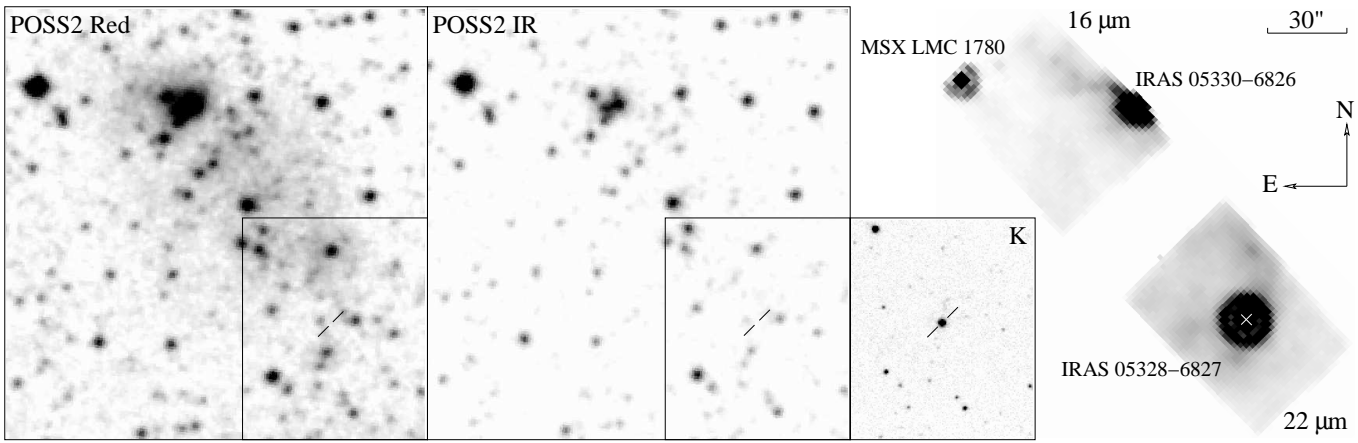,width=180mm}}
\caption[]{A POSS2 red image (left), which includes H$\alpha$ emission, shows
the H\,{\sc ii} region LHA\,120-N\,148\,A and a compact cluster embedded in
it. The cluster stands out better on the POSS2 IR image (centre).
IRAS\,05328$-$6827 only becomes visible at wavelengths beyond $\sim1$ $\mu$m
and is one of the brighter stars in the MSSSO $K$-band image. The IRS red
peak-up image (right) shows an intense source at 22 $\mu$m. Coincidentally,
the blue peak-up image at 16 $\mu$m includes the cluster, IRAS\,05330$-$6826,
and a stellar mid-IR source, MSX\,LMC\,1780.}
\end{figure*}

IRAS\,05328$-$6827 is seen against N\,148\,A, the north-eastern part of the
H\,{\sc ii} region LHA\,120-N\,148 (Fig.\ 1). Compared to 30 Doradus, located
almost a degree to the south-east, and the H\,{\sc ii} region LHA 120-N\,44,
to the north-west, N\,148 is not very bright but it does extend almost half a
degree in diameter. Optical images are suggestive of the presence of a large
dark cloud immediately to the north. N\,148\,A seems to be disjoint from the
rest of N\,148 (Fig.\ 2). There is no evidence for severe interstellar
reddening of stars in the nebula.

IRAS\,05328$-$6827 is a bright object in the $K$-band and in the IRS red
peak-up image at 22 $\mu$m (shown in Fig.\ 2 in vertical alignment with, and
at an identical spatial scale as the optical and $K$-band images). The blue
peak-up image at 16 $\mu$m (shown in Fig.\ 2 in its true position relative to
the red peak-up image) happened to catch the two other bright mid-IR sources
in this region, IRAS\,05330$-$6826 and MSX\,LMC\,1780 (both are easily visible
in Fig.\ 1). The nature of MSX\,LMC\,1780 is unclear, but IRAS\,05330$-$6826
coincides with an anonymous cluster of stars embedded within the H$\alpha$
nebulosity. The mid-IR emission might arise from circumstellar material of
cluster pre-main sequence stars or from diffuse dust that is heated up by the
young stars.

Neither $^{12}$CO$_{1-0}$ (Israel et al.\ 1993) nor radio emission (Marx,
Dickey \& Mebold 1997) is detected from IRAS\,05328$-$6827. The lack of radio
emission (at a level of 6 mJy at 1.4 GHz or 3 mJy at 2.4 GHz) rules out the
possibility that IRAS\,05328$-$6827 is an ultra-compact H\,{\sc ii} region
(cf.\ Zijlstra 1990) unless it is optically thick at those wavelengths.
Interestingly, very weak CO and radio emission at 1.4 GHz is detected from
IRAS\,05330$-$6826 .

\subsection{Modelling the spectral energy distribution}

The spectral energy distribution (SED) was reproduced with the dust radiative
transfer model {\sc dusty} (Ivezi\'{c}, Nenkova \& Elitzur 1999), as a first
attempt at deriving quantitative information about this YSO. The fit is shown
along with the observed SED in Fig.\ 3.

We placed a blackbody with an effective temperature of 30,000 K in the centre
of a spherically symmetric dust cloud. For a distance to the LMC of 50 kpc the
integrated SED yields a bolometric luminosity of 40,000 L$_\odot$. This would
correspond to a late-O type, 20 M$_\odot$ zero-age main-sequence star (Hanson,
Howarth \& Conti 1997). Although the model does not fit the SED in perfect
detail, the bolometric luminosity is recovered with about 10 per cent
accuracy.

For the radial profile of the density in the dust cloud we adopted a single
power law of the form $\rho(r) \propto r^{-0.8}$. Such a flat profile is
consistent with an accreting YSO (Osorio, Lizano \& D'Alessio 1999) or with a
flattened geometry much alike that of a disc (e.g., Whitney et al.\ 2003). The
dust cloud extends from an inner radius of a few hundred AU, at which the dust
temperature reaches 800 K, to an outer radius of about one parsec. The model
fit uses oxygen-rich cold silicates (Ossenkopf, Henning \& Mathis 1992) with a
uniform grain size of 0.07 $\mu$m, and requires an extinction of $A_{\rm
V}=15$ mag.

An improved fit of a spherically symmetric model to the SED cannot be achieved
without invoking a two-component model, with a warm inner dust shell and a
cold outer envelope. However, the geometry in massive YSOs is almost certainly
not spherically symmetric. The present data are unable to constrain detailed
models.

%
%
\begin{figure}
\centerline{\psfig{figure=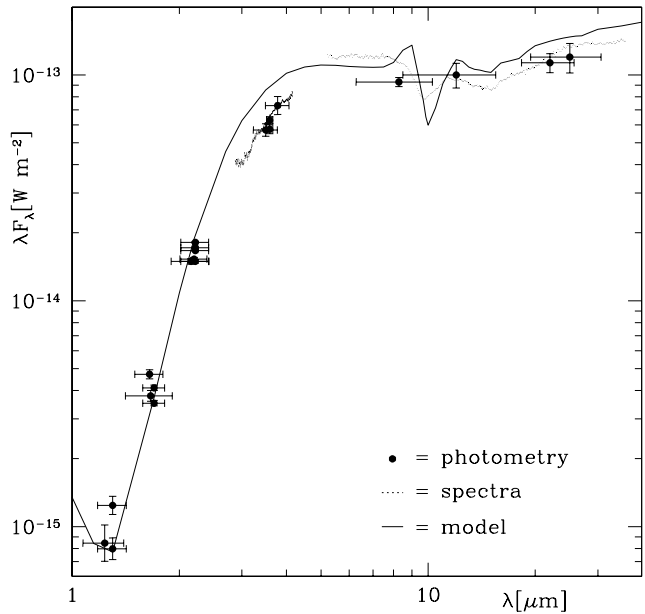,width=84mm}}
\caption[]{Spectral energy distribution of IRAS\,05328$-$6827 as observed via
photometry and spectroscopy between 1 and 36 $\mu$m, and the {\sc dusty} model
fit (see text).}
\end{figure}

\subsection{Spectral analysis: ionization, ices and dust}

The 2.9--4.1 $\mu$m VLT/ISAAC spectrum and 5.2--36 $\mu$m Spitzer/IRS spectrum
are displayed in Fig.\ 4. We define {\it ad hoc} continua (dotted lines) and
compute the optical depths with respect to this pseudo-continuum (Fig.\ 5).
Although this has little meaning for the dust features it facilitates
computing column densities for the molecular ices.

The absence of any atomic emission lines (Fig.\ 4) makes it unlikely that an
advanced ionized region is present. We estimate upper limits to individual
hydrogen and helium emission line fluxes of $F_{\rm line}<1\times10^{-17}$ W
m$^{-2}$, which corresponds to a luminosity of $L_{\rm line}<3\times10^{26}$ W
(=1.5 L$_\odot$) or $L_{\rm photon}<6\times10^{45}$ s$^{-1}$. In a Case B
scenario, in which all lines except Ly$\alpha$ are optically thin, the
strongest line in the 2.9--4.1 $\mu$m region is expected to be Br$\alpha$. The
Br$\alpha$:Br$\gamma$ intensity ratio is $\sim3$ (Storey \& Hummer 1995) and
by scaling we obtain an upper limit of $10^{26}$ W to the Br$\gamma$
luminosity. This is a typical value for detected Br$\gamma$ in galactic
massive YSOs but it is not unusual for it to be weaker (Ishii et al.\ 2001).

The heavily smoothed VLT/ISAAC spectrum (Fig.\ 5) shows absorption by H$_2$O
(water) ice around 3.1 $\mu$m. There is also evidence for absorption by
CH$_3$OH (methanol) ice at 3.4 and 3.5 $\mu$m. The 3.4 $\mu$m absorption in
galactic YSOs often peaks at 3.47 $\mu$m, but a peak around 3.40 $\mu$m as
seen in IRAS\,05328$-$6827 is predicted if relatively little H$_2$O and
H$_2$CO (formaldehyde) ice is present (Schutte et al.\ 1996). Related features
might be detected at 3.85 and 3.95 $\mu$m (Dartois et al.\ 1999). The optical
depth ratio at 3.1 and 3.47 $\mu$m is with $\sim3$ an order of magnitude
smaller than observed in galactic YSOs (Brooke, Sellgren \& Geballe 1999).

There is clear evidence for absorption at 15.2 $\mu$m due to CO$_2$ ice (Fig.\
5); both the position and shape are identical to that seen in the galactic
massive YSO W\,33\,A (Boogert et al.\ 2004). CO$_2$ ice is expected in a
low-mass embedded YSO (Bergin et al.\ 2005), but it is often relatively weak
in {\it massive} YSOs (Watson et al.\ 2004) unless significant processing has
occurred (Gibb et al.\ 2004).

The integrated optical depths $\int\tau{\rm d}\nu$ (wavenumber $\nu$) of the
ice features are converted into column densities by division through by the
integrated absorbance values from Hudgins et al.\ (1993). We thus obtain
$N($H$_2$O$)\simeq1.5\times10^{17}$ cm$^{-2}$, and relative abundances of
approximately H$_2$O:CH$_3$OH:CO$_2$=2:3:1. Although these abundances come
with considerable uncertainty, CH$_3$OH seems to be more abundant than CO$_2$
and H$_2$O is clearly under-abundant. The galactic massive YSOs W\,33\,A and
AFGL\,7009\,S also have an exceptionally high CH$_3$OH abundance, but only
marginally higher than the CO$_2$ abundance and a few times below that of
H$_2$O (Dartois et al.\ 1999). IRAS\,05328$-$6827 may thus present an extreme
case of processing of the ice mantles on the dust grains (cf.\ Gibb et al.\
2004).

We speculate that the low metallicity in the LMC may lead to a less tenuous
dust environment surrounding YSOs, allowing interstellar photons to penetrate
deeper and to be scattered around within the extended envelope. The photons
can then affect the ice coatings on the grain surfaces. The lower oxygen
abundance and possibly higher C/O ratio in the ISM of the LMC may also have an
effect on the chemical balance in YSOs.

The observed opacity ratio of the 18 and 10 $\mu$m silicate components
exceeds any that could be reproduced using the optical constants available to
us. This ratio is known to be high at the low temperatures encountered in the
ISM and star forming regions, as compared to the warm circumstellar envelopes
of evolved stars (Ossenkopf et al.\ 1992; Suh 1999). It goes beyond the
present analysis to design a grain with matching optical properties.

%
%
\begin{figure}
\centerline{\psfig{figure=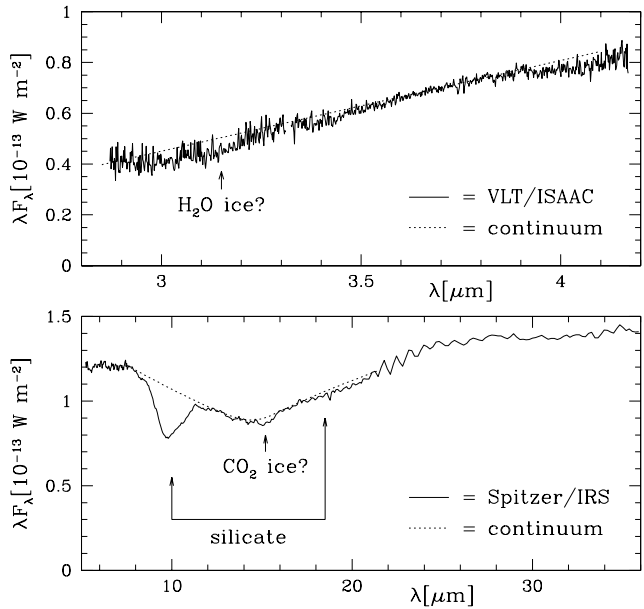,width=84mm}}
\caption[]{VLT/ISAAC (top) and Spitzer/IRS (bottom) spectra of
IRAS\,05328$-$6827. {\it Ad hoc} continua (dotted) are drawn to aid in
analysing the absorption features (see Fig.\ 5).}
\end{figure}

%
%
\begin{figure}
\centerline{\psfig{figure=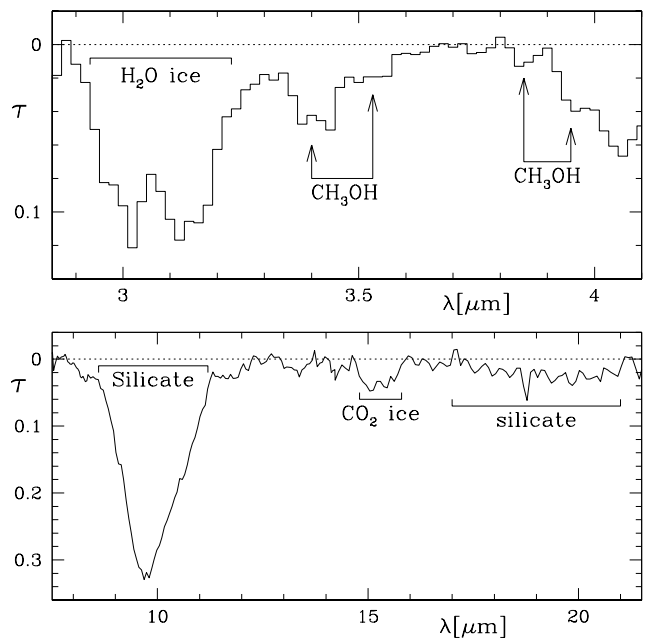,width=84mm}}
\caption[]{VLT/ISAAC (top, heavily smoothed) and Spitzer/IRS (bottom) spectra
of IRAS\,05328$-$6827, expressed in optical depth with respect to the
continuum drawn in Fig.\ 4. There is clear evidence for absorption by ices of
H$_2$O, CH$_3$OH and CO$_2$.}
\end{figure}

\section{Conclusions}

We have discovered a massive Young Stellar Object, IRAS\,05328$-$6827, in the
LMC H\,{\sc ii} region LHA\,120-N\,148. We present spectra at 2.9--4.1 $\mu$m
obtained with the ESO-VLT and at 5.2--36 $\mu$m obtained with the Spitzer
Space Telescope: the first thermal-IR spectra of an extra-galactic YSO. This
opens the way to studying the early stages of massive star formation under
unique environmental conditions (e.g., at low metallicity) not encountered in
the Milky Way.

IRAS\,05328$-$6827 is a luminous ($40,000$ L$_\odot$), massive (20 M$_\odot$)
YSO embedded in an optically thick dust cloud. Its SED is reminiscent of Class
I or flat-spectrum sources, and shows absorption features of silicate dust and
ices. There is no sign yet of an emerging H\,{\sc ii} region.

At the lower metallicity in the LMC, compared to star forming regions in the
Milky Way (by about a factor 4, Koornneef 1982), the lower dust content has
two effects: (i) it becomes easier for the observer to detect the inner,
accreting part of the cloud core, and (ii) interstellar radiation can
penetrate deeper into the cloud and affect the ice mantles on the dust grains.
Together with a higher carbon:oxygen ratio this might be responsible for the
observed low H$_2$O ice abundance and high CH$_3$OH abundance.

\section*{Acknowledgments}

We thank the referee for her/his very helpful suggestions. JO and MM
acknowledge support from PPARC. PRW has received funding for this work from
the Australian Research Council. The research presented here is based on data
collected at the European Southern Observatory (72.D-0351) and Spitzer Space
Telescope (GO 3505), and has made use of the 2MASS and MSX databases and of
Simbad.

\label{lastpage}


\begin{thebibliography}{}
\bibitem[]{BerginEtal2005}
Bergin E.A., Melnick G.J., Gerakines P.A., Neufeld D.A., Whittet D.C.B.\ 2005,
ApJ 627, L33
\bibitem[]{BoogertEtal2004}
Boogert A.C.A., Pontoppidan K.M., Lahuis F., et al.\ 2004, ApJS 154, 359
\bibitem[]{BrandnerEtal2001}
Brandner W., Grebel E.K., Barb\'{a} R.H., Walborn N.R., Moneti A.\ 2001, AJ
122, 858
\bibitem[]{BrookeSellgrenGeballe1999}
Brooke T.Y., Sellgren K., Geballe T.R.\ 1999, ApJ 517, 883
\bibitem[]{CohenWheatonMegeath2003}
Cohen M., Wheaton Wm.A., Megeath S.T.\ 2003, AJ 126, 1090
\bibitem[]{DartoisEtal}
Dartois E., Schutte W., Geballe T.R., Demyk K., Ehrenfreund P., d'Hendecourt
L.\ 1999, A\&A 342, L32
\bibitem[]{EganPrice1996}
Egan M.P., Price S.D.\ 1996, AJ 112, 2862
\bibitem[]{EganVandykPrice2001}
Egan M.P., Van Dyk S.D., Price S.D.\ 2001, AJ 122, 1844
\bibitem[]{EpchteinBrazSevre1984}
Epchtein N., Braz M.A., S\`{e}vre F.\ 1984, A\&A 140, 67
\bibitem[]{GarayLizano1999}
Garay G., Lizano S.\ 1999, PASP 111, 1049
\bibitem[]{GibbEtal2004}
Gibb E.L., Whittet D.C.B., Boogert A.C.A., Tielens A.G.G.M.\ 2004, ApJS 151,
35
\bibitem[]{HansonHowarthConti1997}
Hanson M.M., Howarth I.D., Conti P.S.\ 1997, ApJ 489, 698
\bibitem[]{HouckEtal2004}
Houck J.R., Roellig T.L., van Cleve J., et al.\ 2004, ApJS 154, 18
\bibitem[]{HudginsEtal1993}
Hudgins D.M., Sandford S.A., Allamandola L.J., Tielens A.G.G.M.\ 1993, ApJS
86, 713
\bibitem[]{IshiiEtal2001}
Ishii M., Nagata T., Sato S., Yao Y., Jiang Z., Nakaya H.\ 2001, AJ 121, 3191
\bibitem[]{IsraelEtal1993}
Israel F.P., Johansson L.E.B., Lequeux J., et al.\ 1993, A\&A 276, 25
\bibitem[]{IvezicNenkovaElitzur1999}
Ivezi\'{c} \v{Z}., Nenkova M., Elitzur M.\ 1999, User manual for {\sc dusty}.
University of Kentucky Internal Report
\bibitem[]{Koornneef1982}
Koornneef J.\ 1982, A\&A 107, 247
\bibitem[]{MaerckerBurton2005}
Maercker M., Burton M.G.\ 2005, A\&A 438, 663
\bibitem[]{MarxDickeyMebold1997}
Marx M., Dickey J.M., Mebold U.\ 1997, A\&AS 126, 325
\bibitem[]{Mcgregor1994}
McGregor P.J.\ 1994, PASP 106, 508
\bibitem[]{OsorioLizanoDalession1999}
Osorio M., Lizano S., D'Alessio P.\ 1999, ApJ 525, 808
\bibitem[]{OssenkopfHenningMathis1992}
Ossenkopf V., Henning Th., Mathis J.S.\ 1992, A\&A 261, 567
\bibitem[]{RomanielloRobbertoPanagia2004}
Romaniello M., Robberto M., Panagia N.\ 2004, ApJ 608, 220
\bibitem[]{RubioEtal1998}
Rubio M., Barb\'{a} R.H., Walborn N.R., Probst R.G., Garc\'{\i}a J., Roth
M.R.\ 1998, AJ 116, 1708
\bibitem[]{SchutteEtal1996}
Schutte W.A., Gerakines P.A., Geballe T.R., van Dishoeck E.F., Greenberg J.M.\
1996, A\&A 309, 633
\bibitem[]{SloanEtal2004}
Sloan G.C., Charmandaris V., Fajardo-Acosta S.B., et al.\ 2004, ApJ 614, L77
\bibitem[]{StoreyHummer1995}
Storey P.J., Hummer D.G.\ 1995, MNRAS 272, 41
\bibitem[]{Suh1999}
Suh K.-W.\ 1999, MNRAS 304, 389
\bibitem[]{Vanloon2000}
van Loon J.Th.\ 2000, A\&A 354, 125
\bibitem[]{VanloonZijlstra2001}
van Loon J.Th., Zijlstra A.A. 2001, ApJ 547, L61
\bibitem[]{WalbornEtal1999}
Walborn N.R., Barb\'{a} R.H., Brandner W., Rubio M., Grebel E.K., Probst R.G.\
1999, AJ 117, 225
\bibitem[]{Walmsley1995}
Walmsley C.M.\ 1995, Rev.\ Mex.\ Ast.\ Ap.\ Conf.\ Ser.\ 1, 137
\bibitem[]{WatsonEtal2004}
Watson D.M., Kemper F., Calvet N., et al.\ 2004, ApJS 154, 391
\bibitem[]{WernerEtal2004}
Werner M.W., Roellig T.L., Low F.J., et al.\ 2004, ApJS 154, 1
\bibitem[]{WhiteoakEtal1983}
Whiteoak J.B., Wellington K.J., Jauncey D.L., Gardner F.F., Forster J.R.,
Caswell J.L., Batchelor R.A.\ 1983, MNRAS 205, 275
\bibitem[]{WhitneyEtal2003}
Whitney B.A., Wood K., Bjorkman J.E., Wolff M.J.\ 2003, ApJ 591, 1049
\bibitem[]{WoodHabingMcgregor1998}
Wood P.R., Habing H.J., McGregor P.J.\ 1998, A\&A 336, 925
\bibitem[]{Zijlstra1990}
Zijlstra A.A.\ 1990, A\&A 234, 387
\end{thebibliography}
\end{document}